\documentstyle[preprint,prb,aps]{revtex}
\begin{document}
\draft
\input{psfig}
\title{Quantum Fluctuations in the ANNNI Model}
\author{A. B. Harris,$^{1,2}$ C. Micheletti,$^1$ and J. M. Yeomans$^1$}
\address{ (1) Theoretical Physics, Oxford University,
1 Keble Rd. Oxford OX1 3NP, UK}
\address{(2) Department of Physics,
University of Pennsylvania, Philadelphia, PA 19104-6396}
\date{\today}
\maketitle
\begin{abstract}
We obtain the ANNNI model from a Heisenberg model with
large single--ion anisotropy energy, $D$, as might be
relevant for helical spin systems.  We treat quantum
fluctuations to lowest order in $1/S$ at zero temperature
within an expansion in $J/D$, where $J$ is an exchange energy.
The transition from the state with periodicity $p=4$ to the
uniform state ($p= \infty$) occurs via an infinite sequence of
first order transitions in which $p$ increases monotonically.
\end{abstract}


Systems with long--period modulated structures are surprisingly
common in nature.  Examples include helical phases in the
rare--earths and their compounds[\onlinecite{JJ}],
polytypism[\onlinecite{PT}], and
the arrangement of antiphase boundaries in binary
alloys[\onlinecite{BA}].
A given compound may exhibit many different modulated structures
of differing wavelength as a control parameter such as the
temperature is varied.  Some modulated structures can usefully
be viewed as an assembly of domain walls when the energy for
introducing a wall passes through zero.  The stability of the
different structures is then determined by the interactions
between pairs, trios, etc. of walls[\onlinecite{AMSMEF}].
It has been established that these interactions can result
from entropic contributions to the free energy[\onlinecite{MEFWS}]
and from softening of the spins[\onlinecite{JY}].
Here our aim is to show that quantum fluctuations can also
stabilize long--period modulated structures.

The Hamiltonian we consider is
\begin{eqnarray}
\label{HAMIL}
{\cal H} =
&& - {J_1 \over S^2} \sum_{i,j} {\bf S}_{i,j} \cdot {\bf S}_{i+1,j}
+ {J_2 \over S^2} \sum_{i,j} {\bf S}_{i,j} \cdot {\bf S}_{i+2,j}
\nonumber \\ &&
- {J_0 \over S^2} \sum_{i \langle jj'\rangle} {\bf S}_{i,j}
\cdot {\bf S}_{i,j'}  -  {D \over S^2} \sum_{i,j} ([S_{i,j}^z]^2 -S^2)\ ,
\end{eqnarray}
where $i$ labels the planes of a cubic lattice perpendicular to
the $z$-direction and $j$ the position within the plane.  Also
$\langle jj'\rangle$ indicates a sum over pairs of nearest
neighbors in the same plane and ${\bf S}_{i,j}$ is a quantum
spin of magnitude $S$ at site $(i,j)$.  For $D= \infty$, only
the states $S_{iz}=\sigma_iS$, where $\sigma_i = \pm 1$ are
relevant and ${\cal H}$ reduces to the axial next-nearest
neighbor Ising (ANNNI) model, first proposed to describe
helical phases of the heavy rare earths[\onlinecite{RJE}],
\begin{eqnarray}
{\cal H}_A & = &
- J_0 \sum_{i \langle jj'\rangle} \sigma_{i,j}
\sigma_{i,j'} - J_1 \sum_{i,j} \sigma_{i,j} \sigma_{i+1,j}
\nonumber \\ && \
+ J_2 \sum_{i,j} \sigma_{i,j} \sigma_{i+2,j} \ .
\end{eqnarray}
The ground state of the ANNNI model is ferromagnetic for
$\kappa \equiv J_2/J_1 < 1/2$ and an antiphase structure with
layers ordering in the sequence $\{ \sigma_i \} = \{ \dots
1, 1, -1 , -1, 1,1,-1, -1\dots \}$ for $\kappa > 1/2$.
$\kappa=1/2$ is a multiphase point[\onlinecite{MEFWS}],
where the ground state is
infinitely degenerate with all possible configurations of
ferromagnetic and antiphase orderings having equal energy.
For classical spins $S=\infty$, the ground state (and therefore
the multiphase point) is maintained as $D$ is reduced from
infinity as long as $D$ is larger than about 1/2.  For
higher order anisotropies this is not the case[\onlinecite{JY}].

To describe how the degeneracy is broken at the multiphase
point we use a notation similar to that of
Fisher and Selke[\onlinecite{MEFWS}] so that
$\langle n_1, n_2, \dots n_m \rangle$ denotes a state in
which spins form domains (of parallel spins) whose widths
repeat periodically the sequence $n_1, n_2, \dots n_m$.

Fisher and Selke[\onlinecite{MEFWS}]
showed that at nonzero temperature $T$ the degeneracy at the
multiphase point is broken to give a sequence of phases
$\langle 2^k 3 \rangle$, for $k=1, 2, 3 \dots$.  Fisher and
Szpilka[\onlinecite{AMSMEF,MEFXS}]
later recast their analysis in terms of domain
wall interactions and we will follow their formulation.

In view of this interesting phase diagram in the
$\kappa$-$T$ plane, we are led to study the phase diagram in
the $\kappa$--$D^{-1}$ plane when the spins are quantum operators.
That quantum fluctuations can remove ground--state degeneracies
was pointed out by Shender[\onlinecite{EFS}] and given the apt name
"ground state selection" by Henley[\onlinecite{CLH}].
In this paper we show how the multiphase
degeneracy is resolved by quantum fluctuations.

To study quantum fluctuations we introduce the
Dyson-Maleev[\onlinecite{DM}] transformation
\begin{eqnarray}
S_i^z & = & \sigma_i ( S - a_i^+ a_i) \nonumber \\
S_i^+ & = & \sqrt{2S} \left(
\delta_{\sigma_i,1} \left[ 1 - {a_i^+ a_i \over 2S} \right] a_i +
\delta_{\sigma_i,-1} a_i^+ \left[ 1-{a_i^+ a_i\over 2S} \right]
\right) \nonumber \\
S_i^- & = & \sqrt{2S} \left( \delta_{\sigma_i,1} a_i^+
+ \delta_{\sigma_i,-1} a_i \right) \ ,
\end{eqnarray}
where $\delta_{a,b}$ is unity if $a=b$ and is zero otherwise
and $a_i^+$ ($a_i$) creates (destroys) a spin excitation at
site $i$.  We thereby transform the Hamiltonian of Eq. (\ref{HAMIL})
into the bosonic form
\begin{equation}
\label{HAM}
{\cal H} ( \{ \sigma_i \} ) = E_0 + {\cal H}_0
+ V_{||} + V_{\not{\parallel}} + V^{(4)} \ ,
\end{equation}
where $E_0 \equiv {\cal H}_A$,
\begin{eqnarray}
{\cal H}_0 = \sum_{i,j} \Biggl[ && 2\tilde{D} + J_1 \sigma_{i,j} (
\sigma_{i-1,j} +
\sigma_{i+1,j} ) \nonumber \\
&& - J_2 \sigma_{i,j} ( \sigma_{i-2,j} + \sigma_{i+2,j} ) \Biggr]
S^{-1} a_{i,j}^+ a_{i,j}
\end{eqnarray}
with $\tilde{D}=D+2 J_{0}$
and $V_{||}$ ($V_{\not{\parallel}}$) is the interactions between spins
which are parallel (antiparallel)
\begin{eqnarray}
V_{||} = && {1 \over S} \sum_{i,j} \Biggl[ - J_1
X(i,i+1;j)
(a_{i,j}^+ a_{i+1,j} + a_{i+1,j}^+ a_{i,j} )
\nonumber \\ && + J_2
X(i,i+2;j)
(a_{i,j}^+ a_{i+2,j} + a_{i+2,j}^+ a_{i,j} ) \Biggr]
\end{eqnarray}
\begin{eqnarray}
V_{\not{\parallel}} = && {1 \over S} \sum_{i,j} \Biggl[ - J_1
Y(i,i+1;j)
(a_{i,j}^+ a_{i+1,j}^+ + a_{i+1,j} a_{i,j} )
\nonumber \\ && + J_2
Y(i,i+2;j)
(a_{i,j}^+ a_{i+2,j}^+ + a_{i+2,j} a_{i,j} ) \Biggr] \ ,
\end{eqnarray}
where $X(i,i';j)$ [$Y(i,i';j$)] is unity if spins $(i,j)$ and
$(i',j)$ are parallel [antiparallel] and is zero otherwise.  In
Eq. (\ref{HAM}) $V^{(4)}$ represents the four operator terms
proportional to $1/S^2$.  Fluctuations out of the classical ground
state (the boson vacuum) only occur at the walls due to
$V_{\not{\parallel}}$.  We do not consider quantum fluctuations
within a plane, since the phase diagram is determined by the
interplanar quantum couplings.  Also, since the
walls in this three dimensional system are flat at $T=0$,
we may characterize states of the system in terms of distances
between walls.

We now consider the structure of perturbation theory for all states
which are degenerate at the multiphase point $\kappa = 1/2$.
Perturbation theory generates corrections to the diagonal energy of
the classical states in powers of $1/S$ and $J/\tilde{D}$, where $J=J_1$
or $J_2$.  Off--diagonal matrix elements (for example, in which
two domain walls both move through one lattice constant) first
occur in $2S$th order perturbation theory and may be ignored.
We will only include effects of the quadratic Hamiltonian,
i.e. we will work to leading order in $1/S$.

Instead of a direct evaluation of the energy of all
possible phases, we follow the methods of
Fisher and Szpilka[\onlinecite{MEFXS}] and
study the sequence of wall interaction energies: $E_w$, the
energy of an isolated wall; $V_2(n)$, the interaction
energy of two walls separated by $n$ sites; and generally
$V_k(n_1 , n_2 , \dots n_{k-1})$, the interaction energy of $k$
walls with successive separations $n_1$, $n_2$, ... $n_{k-1}$.
In terms of these quantities one may write the total energy
of the system when there are walls at positions $m_i$ as
\begin{eqnarray}
E = && E_0 + n_w E_w + \sum_i V_2(m_{i+1}-m_i) \nonumber \\
& + & \sum_i
V_3(m_{i+2}-m_{i+1},m_{i+1}-m_i)\nonumber \\ & + & \sum_i
V_4(m_{i+3}-m_{i+2},m_{i+2}-m_{i+1},m_{i+1}-m_i) \nonumber \\
& + &  \dots \ ,
\end{eqnarray}
where $E_0$ is the energy with no walls present and $n_w$ is the
number of walls.  The scheme of Ref. [\onlinecite{MEFXS}] for
calculating the general wall potentials $V_k$ is illustrated in
Fig. 1.
Let all spins to the left of the first wall have $\sigma_i=\sigma$
and those to the right of the last wall have $\sigma_i=\eta$ for
$k$ even and $\sigma_i=-\eta$ for $k$ odd.  The energy of such a
configuration is denoted $E_k(\sigma, \eta)$.  If $\sigma=-1$
($\eta=-1$) the left (right) wall is absent.  Then the energy
ascribed to the existence of $k$ walls is given by[\onlinecite{RBG}]
\begin{equation}
V_k(n_1, n_2, \dots n_{k-1}) = \sum_{\sigma , \eta = \pm 1}
\sigma \eta E_k (\sigma , \eta) \ .
\end{equation}
Contributions to $E_k$ which are independent of $\sigma$ or $\eta$
do not influence $V_k$.  $E_k(\sigma , \eta)$ is calculated
by developing the energy in powers of the perturbations
$V_{||}$ and $V_{\not{\parallel}}$.  To lowest order in $1/\tilde{D}$,
contributions to $V_k$ can be obtained, for instance, by creating an
excitation at the left wall (using $V_{\not{\parallel}}$) and (for
wall separations $n_1>3$) using $V_{||}$ to hop the excitation
sufficiently near the other wall that one (or more) energy
denominator depends on $\eta$.  Examples of such processes are
shown in Fig. 2.

For instance for the top diagram of Fig. 2, we get
\begin{equation}
E_2(\sigma, \eta) = -  \left( { J_2^2 \over 4\tilde{D} + J_1 + J_2 + \eta
(J_2-J_1)}
\right) \frac{\delta_{\sigma,1}}{S} \ ,
\end{equation}
which gives a contribution to $V_2(2)$ at order $J^3/\tilde{D}^2S$ of
\begin{equation}
\sum_{\sigma , \eta = \pm 1} \sigma \eta E_2 (\sigma , \eta)
= {2J_2^2 (J_2-J_1) \over 16\tilde{D}^2S} \ .
\end{equation}
Collecting all such processes we find the general result
\begin{eqnarray}
\label{VODD}
&& V_2(2n+1) = {16 \tilde{D} \over S} \left( {J_2 \over 4\tilde{D}}
\right)^{2n+1}
\\ \label {VEV} &&
V_2(2n) = { 4n^2 (J_1^2/J_2) - 4J_1 + 8J_2 \over S }
\left( {J_2 \over 4\tilde{D}} \right)^{2n} \ .
\end{eqnarray}
These results may be understood in terms of a correlation length
$\xi \sim [1/ \ln (4\tilde{D}/J_2)]$ which governs wall--wall interactions.

More generally, power counting shows that
\begin{eqnarray}
V_3(2n,2n) & \sim & V_3(2n,2n+1) \sim J (J/\tilde{D})^{4n} \nonumber \\
V_3(2n-1,2n-1) & \sim & V_3(2n-1,2n) \sim J (J/\tilde{D})^{4n-1}
\end{eqnarray}
and $V_k(n_1, n_2, \dots n_{k-1}) \sim J (J/\tilde{D})^x$, where
$x \geq \sum_j n_j -2$.  Second
order perturbation theory yields the result
\begin{equation}
E_w = 2J_1 - 4J_2 - {J_1^2 + 2J_2^2 \over 4\tilde{D}S } +
{\rm O}(J^3/\tilde{D}^2S) \ .
\end{equation}
When $E_w>0$, the ferromagnetic phase is stable.  This happens
for $J_2 < J_c = J_1/2 - (3J_1^2/8\tilde{D}S) \dots$.

We wish to describe the sequence of phases which occur as
$J_2/J_1$ is decreased starting
from $\langle 2 \rangle$ when $J_2/J_1 > 1/2$ and reaching
$\langle \infty \rangle$ when $J_2<J_c$.  As Fisher and Szpilka
show, the phase boundary along which $\langle n \rangle$
and $\langle n+1 \rangle$ have the same energy is given by
\begin{eqnarray}
E_w = && nV_2(n) - (n+1)V_2(n+1) + nV_3(n,n) \nonumber \\
&&  - (n+1) V_3(n+1,n+1) + \dots
\end{eqnarray}
This relation yields a critical value of $J_2$, denoted $J_{nc}$
which can be expressed as $J_{nc} = J_c + \Delta J_2(n)$, where
\begin{equation}
\label{DEL}
\Delta J_2(n) = {n V_2(n) - (n+1) V_2(n+1) \over
\partial E_w / \partial J_2 } + \dots \biggr|_{J_2=J_1/2} \ .
\end{equation}

Thus, to elucidate the topology of the phase diagram, it is not
necessary to know $J_c$ accurately.  For $n$ not too large,
Eqs. (\ref{VODD}), (\ref{VEV}), and (\ref{DEL}) give
$\Delta J_2(n) \sim V_2(n) \sim J(J^2/\tilde{D}^2)^{[n/2]}$, where $[x]$ is
the integer part of $x$.  The tentative conclusion is that one
has successive regions of stability of the phase $\langle n \rangle$,
where $n$ increases as $J_2$ decreases, as shown in Fig. 3.
However, we must check the stability of the phase boundary to mixed
phases of $\langle n \rangle$ and $\langle n+1 \rangle$.

As Fisher and Szpilka show, the condition that this
phase boundary be stable is that $F_n < 0$, where
\begin{equation}
\label{STAB}
F_n \equiv V_3(n,n) -2 V_3(n,n+1) + V_3(n+1,n+1) \ .
\end{equation}
Here the last term is higher order in $1/\tilde{D}$ than the first two
and can be neglected.  All perturbative terms which contribute
at lowest order in $1/\tilde{D}$ to $V_3(n,n+1)$ have their analogs for
$V_3(n,n)$.  By an appropriate grouping of terms one can show
that for $n>2$, $F_n<0$.  Basically this happens because
even order ground-state-to-ground-state terms in
perturbation theory are negative.  The case $n=2$ is special
in that $F_2=0$ at lowest order.  Then it is necessary to
go to the next order, where we find
\begin{eqnarray}
&& V_3(2,2) = {8J_2^2 \over (4\tilde{D})^3S} [ -J_1^2 + 2J_1J_2 - 2J_2^2]
\nonumber \\ &&  \ +
{12J_2^2 \over (4\tilde{D})^4S } [ -4J_1^3 + 12 J_1^2J_2 - 5 J_1J_2^2 + 10
J_2^3 ]
\end{eqnarray}
\begin{equation}
V_3(2,3) = - {8J_2^4 \over (4\tilde{D})^3S} + {12J_2^2 \over (4\tilde{D})^4S} [
2J_1^2 J_2
+ 4 J_1 J_2^2 + 5J_2^3 ] \ .
\end{equation}
\begin{equation}
V_3(3,3) = {\rm O} (J^6/\tilde{D}^5S) \ .
\end{equation}
To leading order in $1/S$ we may set $J_2=J_1/2$, in which case
the above results indicate that $F_2 \sim A/\tilde{D}^4$, where $A<0$.
Thus all the phase boundaries between phases $\langle n \rangle$ and
$\langle n+1 \rangle$ are stable agains subdivision.

The above results are valid (as we shall see)
for $n \ll \sqrt {\tilde{D}/J}$.  When this limit is
violated, the entropy of more complicated perturbation contributions
can compensate for taking more powers of $J/\tilde{D}$.  We overcome this
limitation with respect to $V_2(n)$ as follows.  We work to lowest
(second) order in $V_{\not{\parallel}}$ (A pair of excitations
is created, one to the left of the left wall and one to the right of the
left wall, as in Fig. 2, and is later destroyed.)
To simplify the result we assume that
the excitation created to the left of the left wall does not
propagate.  We work to first order in the field exerted on spins
$n-1$ and $n$ by the spins in the neighboring domain.  The result
for the ground state energy is then expressed in terms of the EXACT
spin-wave Green's function, $G^{(n)}$, for an isolated domain
of $n$ spins.  In this way we sum over all trajectories of the
spin deviation inside the domain of $n$ parallel spins.  For
small $n$ we reproduce the above results[\onlinecite{GN}].
For large $n$ the result at leading order in $J/\tilde{D}$ is
\begin{equation}
\label{VN}
V_2(n) = 4J_2^3 G^{(n)}(2,n-1)^2/S \ ,
\end{equation}
where
\begin{equation}
G^{(n)}(i,j) = \sum_\alpha { \phi_\alpha(i) \phi_\alpha (j)
\over 2\tilde{D} + \epsilon_\alpha } \ .
\end{equation}
Here $\phi_\alpha$ and $\epsilon_\alpha$ are the exact
eigenstates and energies for the single--spin excitations of
an isolated system of $n$ parallel spins.  We carried out an
exact evaluation of $G^{(n)}(2,n-1)$. For large $n$, we found
\begin{eqnarray}
\label{G2}
G(2,n-1)^2 & = & {4\tilde{D} \over J_2^3 } e^{-n / \tilde \xi }
\sin^2 (n \delta ) \ , \ \ n \ \ {\rm even} \nonumber \\ & = &
{4\tilde{D} \over J_2^3 } e^{-n / \tilde \xi }
\cos^2 (n\delta  )\ , \ \ n \ \ {\rm odd} \ ,
\end{eqnarray}
where $\delta=J_1 / \sqrt { 16\tilde{D}J_2}$
and $\tilde \xi - \xi$ differs from zero due to corrections
which are higher order in $J/\tilde{D}$.  Eqs. (\ref{VN}) and
(\ref{G2}) seem to imply that $V_2(n)$ can become arbitrarily
close to zero.  That is an artifact of truncating these
equations at leading order in $J/\tilde{D}$.  Where $V_2(n)$ would
be small, one must keep the appropriate terms which are
otherwise corrections.  So doing we have a result which
is uniformly asymptotically correct for
$n \gg \sqrt {\tilde{D}/J}$[\onlinecite{EX}]:
\begin{eqnarray}
\label{LARGE}
V_2(n) & = & {16\tilde{D} \over S } e^{-n / \tilde \xi }
\biggl( \sin^2 (n \delta + \phi ) + {J_1J_2 \over 2\tilde{D}^2 } \biggr) ,
\ n \ {\rm even} \nonumber \\
& = &
{16\tilde{D} \over S } e^{-n / \tilde \xi } \biggl(
\cos^2 (n \delta   + \phi ) + {J_1J_2 \over 2\tilde{D}^2 } \biggr)
, n \ {\rm odd} , \nonumber \\ &&
\end{eqnarray}
where $\phi$ is a phase shift of order $1/ \sqrt {\tilde {D}}$.

An elegant
graphical interpretation of the phase boundaries suggested by
Fisher and Szpilka is that one should construct the extremal convex
envelope of $V_2(n)$ versus $n$.  The points [$n,V_2(n)$] which
make up the envelope correspond to the phases $\langle n \rangle$
which occur when $V_2(n)$ is not convex, as Eq. (\ref{G2}) shows
to be the case.  As a result, we conclude that there is an infinite
sequence of phase boundaries.  When $2n \delta / \pi$ is nearly
an integer, the phase boundaries will be between phases
$n$ and $n+2$ because of the nonconvexity of $V_2(n)$.  Note
that in contrast to the ANNNI model, here $V_2(n)$ does not pass
through zero.  This difference can be understood as follows.
In the present model in order for an excitation to sense the
presence of a second wall, it has to travel from one wall to
the other wall and return, giving rise to the factor
$G^2$ in Eq. (22).  In the ANNNI model the analogous
factor involves only a one-way connection corresponding to
$G$.  As a consequence, for the
ANNNI model the sequence of phases terminates at a value,
$n_0$, which diverges as $T \rightarrow 0$.  There is no
cut-off on $n$ in the present model.

We were unable to carry out a
precise analysis for $V_3$ at large $n$.
Accordingly, at large $n$ we can not guarantee the stability
of these phase boundaries.  It is
conceivable that our result for small $n$ breaks down and
that the phase boundaries obtained from $V_2(n)$ become
unstable to mixing, which could even be hierarchical.

To summarize:  1) We have shown that quantum fluctuations do
remove the infinite degeneracy of the multiphase point
of the ANNNI model.  2) We have shown that quantum
fluctuations at $T=0$ lead to a sequence of first order
transitions similar to that for the ANNNI model, but
involving a different sequence of phases.  3) In contrast
to the ANNNI model there is no cut-off at large $n$ on the
appearance of phases because here $V_2(n)$ never becomes
negative.  As we explained, this is a peculiarly quantum effect.

ACKNOWLEDGEMENTS:
JMY is supported by an EPSRC Advanced Fellowship,
ABH by an EPSRC Visiting Fellowship, and CM by an
EPSRC Studentship and the Fondazione
"A. della Riccia," Firenze.

\newpage
\begin{center}
{\large\bf FIGURE CAPTIONS}
\end{center}
\vspace{0.5cm}

\noindent
{\bf FIG. 1}
Configurations needed to calculate the interaction
energy for two walls at separation $n$ (top) and three
walls at separations $n$ and $m$ (bottom).

\noindent
{\bf FIG. 2}
Examples of configurations needed to calculate $V_2(2)$ (top),
$V_2(3)$ (middle), and $V_2(4)$ (bottom). Here "+" ("-")
indicate creation (destruction) of a spin excitation and the arrow
indicate a hopping using $V_{||}$.

\noindent
{\bf FIG. 3}
Schematic phase diagram of the "soft" ANNNI model.
The phase boundary between $\langle n \rangle$ and $\langle n+1 \rangle$
depends on a power of $1/\tilde{D}$ which increases with $n$.  We did not
attempt
to represent this dependence on $\tilde{D}$ correctly.

\newpage

\vspace*{2in}

\centerline{\psfig{figure=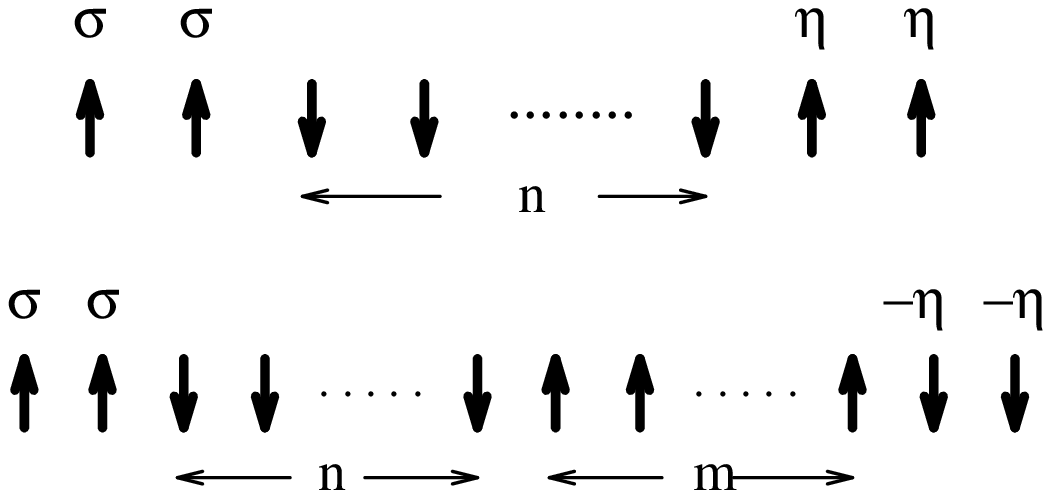}}
\vspace{2cm}
{\bf\Huge  FIG.1}

\newpage

\vspace*{2in}

\centerline{\psfig{figure=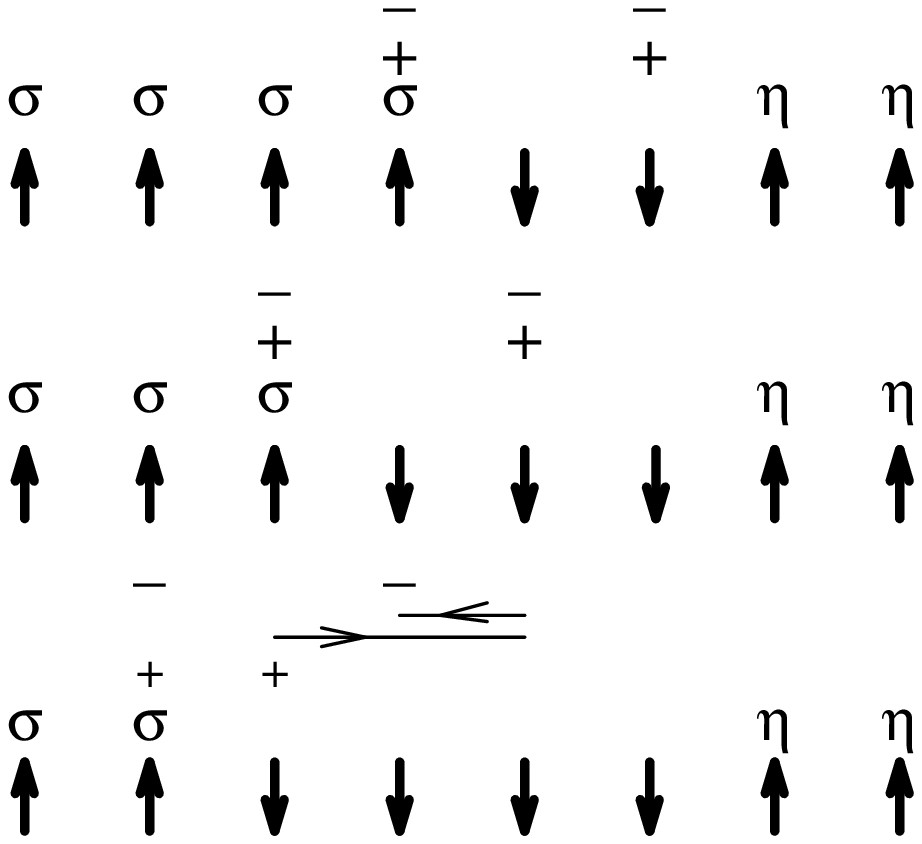}}
\vspace{2cm}
{\bf\Huge  FIG.2}

\newpage

\vspace*{2in}

\centerline{\psfig{figure=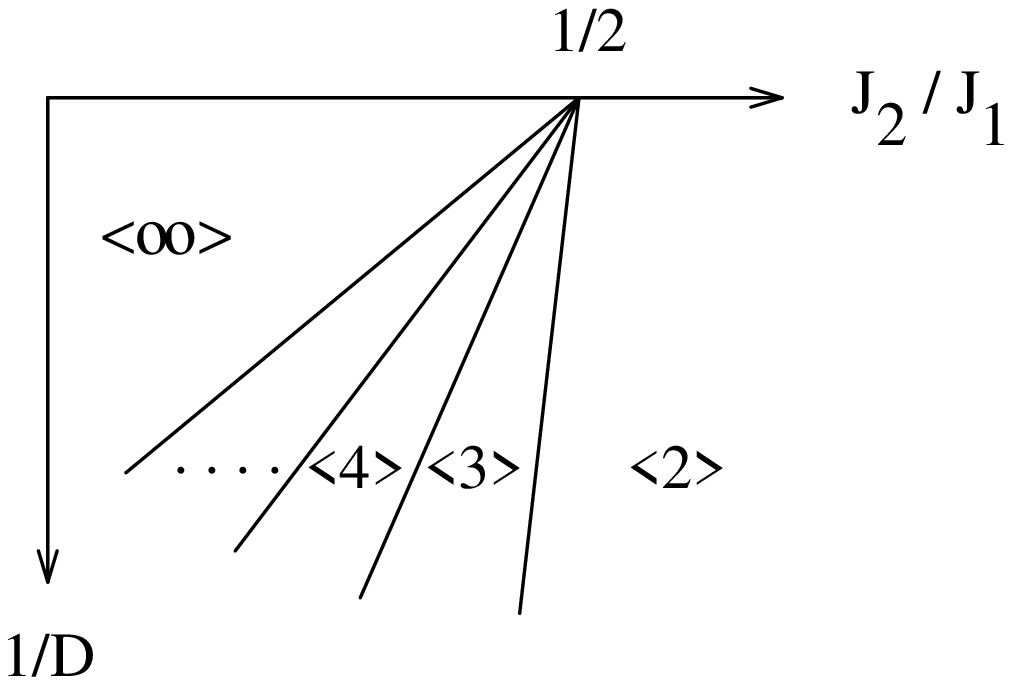}}
\vspace{2cm}
{\bf\Huge  FIG.3}


\begin{references}
\bibitem{GN}
For small $n$ and to lowest order in $J/\tilde{D}$, Eq. (\ref{VN})
holds for $n$ odd.  For $n$ even and small we have
$V_2(n) = 4 J_2^3 [ G^{(n)}(2,n-1) -2 G(1,n-1)]^2/S$
\bibitem{EX}
We also checked numerically (for $\tilde{D}/J_2$ up to 400) that
in the intermediate region (where neither the small $n$ nor
the large $n$ results apply) $V_2(n)$ remains positive.  In
fact, Eq. (\ref{LARGE}) is a reasonable approximation for all $n$.
\end{references}
\end{document}